\documentclass[conference]{IEEEtran}
\IEEEoverridecommandlockouts
\usepackage{cite}
\usepackage{amsmath,amssymb,amsfonts}
\usepackage{algorithmic}
\usepackage{graphicx}
\usepackage{textcomp}
\usepackage{url}
\usepackage{xcolor}
\usepackage[numbers]{natbib} 
\def\BibTeX{{\rm B\kern-.05em{\sc i\kern-.025em b}\kern-.08em
    T\kern-.1667em\lower.7ex\hbox{E}\kern-.125emX}}
    
\begin{document}

\title{System Misuse Detection via Informed Behavior Clustering and Modeling}

\author{\IEEEauthorblockN{Linara Adilova\IEEEauthorrefmark{1}\IEEEauthorrefmark{2}, Livin Natious\IEEEauthorrefmark{1}, Siming Chen\IEEEauthorrefmark{1}\IEEEauthorrefmark{4}, Olivier Thonnard\IEEEauthorrefmark{3}, and Michael Kamp\IEEEauthorrefmark{1}\IEEEauthorrefmark{2}\IEEEauthorrefmark{4}}\\
\IEEEauthorblockA{\IEEEauthorrefmark{1}Fraunhofer Institute for Intelligent Analysis and Information Systems (IAIS), Germany\\
Email: \{linara.adilova, livin.natious.livin.natious, siming.chen, michael.kamp\}@iais.fraunhofer.de}
\IEEEauthorblockA{\IEEEauthorrefmark{2}Fraunhofer Center for Machine Learning, Germany\\
}
\IEEEauthorblockA{\IEEEauthorrefmark{3}Amadeus, France\\
Email: olivier.thonnard@amadeus.com
}
\IEEEauthorblockA{\IEEEauthorrefmark{4}University of Bonn, Germany
}
\thanks{This research is supported by EU Project DiSIEM (Grant no. 700692).}

}


\maketitle

\begin{abstract}
    One of the main tasks of cybersecurity is recognizing malicious interactions with an arbitrary system. Currently, the logging information from each interaction can be collected in almost unrestricted amounts, but identification of attacks requires a lot of effort and time of security experts. We propose an approach for identifying fraud activity through modeling normal behavior in interactions with a system via machine learning methods, in particular LSTM neural networks. In order to enrich the modeling with system specific knowledge, we propose to use an interactive visual interface that allows security experts to identify semantically meaningful clusters of interactions. These clusters incorporate domain knowledge and lead to more precise behavior modeling via informed machine learning. We evaluate the proposed approach on a dataset containing logs of interactions with an administrative interface of login and security server. Our empirical results indicate that the informed modeling is capable of capturing normal behavior, which can then be used to detect abnormal behavior. 
\end{abstract}

\section{Introduction}
\label{introduction}
In this paper we consider the context of a system misuses identification. We assume that the system allows for a fixed set of actions, e.g., system calls in any operating system or actions like 'SearchItem', 'FilterResults' in any online shop. Logs of interactions with a system can be used as data for training automated security models for protecting it from intrusions. Usually, the analysis of logs is performed manually by security specialists, since the usage of automated methods is prone to many possible problems. One of the prominent ones is the regulation of false alarms--usually it is hard to generalize for detection of unseen attacks because they are mostly mimicking normal behavior \citep{sommer2010outside}. Even more, intrusion detection is considered to be a hard task for automated methods because of particular challenges in the field \citep{sommer2010outside}: 
\begin{itemize}
    \item semantic gap between the predictions and interpretable information makes it hard to use the model outputs by security experts;
    \item modeling normal traffic without becoming too vague is a challenge due to the diversity of network traffic to be analyzed;
    \item evaluation schemes are frequently not capable of capturing true properties of a model.
\end{itemize}

Nevertheless, novel machine learning approaches are promising to achieve good results on the task of intrusion detection. In particular, classical approaches were applied, e.g., SVMs and decision trees; artificial neural networks were also considered in more recent works \cite{xin2018machine}. There exist two main classes of intrusion detection approaches:
\begin{enumerate}
    \item based on the known attack schemes and their recognition in system interactions;
    \item based on anomaly detection via learning normal behavior in the system and finding outlying behavior.
\end{enumerate}
The first class of approaches is the most precise methods for detecting existing attacks if the application environment is restricted to a specific system with known vulnerabilities. The main problem with such methods is a need of constant support of the databases containing attack signatures. The more popular class for applying machine learning methods is the second one, requiring generalization to unseen cases and allowing more freedom in application systems.

Since machine learning approaches are more suitable for learning previously seen normal behavior than recognizing unseen malicious ones \citep{sommer2010outside}, a promising technique is to learn behavioral patterns from the activity in the system in a continuous way. Then a fraud can be detected as an outlier by monitoring the current activities using learned models for normal behavior. The main challenge in such application is the variety of possible activities that requires very thorough and long observation of the system interactions. The proposed solution is to subdivide the activities according to some criterion. For example, \citet{chandrasekhar2014confederation} employed fuzzy clustering to separate data into homogeneous subsets and learn separate artificial neural networks for each of the subsets for further predictions. The best option for consistency of modeled behaviors is to subdivide activities according to semantically meaningful properties which requires domain and system specific knowledge.

The importance of target systems understanding in such specific domain is quite high \citep{sommer2010outside}. It is critical for the detection of known attacks, but can also be used for validating and evaluating normal behavior and discovering anomalies through it. This leads to a necessity of informed machine learning techniques. For example, \citet{meng2015design} involved experts knowledge in the loop for recognizing the most successful models for reducing the amount of false alarms, at the same time keeping the level of accuracy high.

Overall, \citet{xin2018machine} in their survey indicate that machine learning and deep learning approaches can be quite successfully applied in cybersecurity and show high performance. 
The most successful models are recurrent neural networks, in particular Long Short-Term Memory (LSTM) networks \citep{kim2016lstm}.

In this paper we consider the task of identifying misuses of an administrative interface of an internal portal. The use case is identified during the DiSIEM project\footnote{\url{http://www.disiem-project.eu/}}. Because of the severity of the actions inside the portal it is crucial for operators to identify misuses, e.g., sessions indicating abuse of personal information, such as deleting ('ActionDeleteUser') or resetting access ('ActionResetPwdUnlock') of multiple user profiles \citep{nguyen2018understanding}. To allow monitoring, the interactions are logged as sessions containing sequences of actions. The security operators regularly check all the activity in order to identify possibly malicious behavior. The recorded dataset consists of the recorded activities of normal behavior and thus, the main task is to model such normal behavior. Automated identification of normality will reduce operators work by allowing them to only investigate suspicious interactions. We propose using LSTM-based language models for learning normal behavior for clusters of interactions identified through a visual system by security experts. All new interactions will be monitored in realtime to identify how much they are aligned with the modeled normal behavior.

The paper is organized in the following parts: first we provide related work, then describe our approach, afterwards we show the results of its evaluation, and conclude with a discussion and future work.
\section{Related Work}
\label{related}
As it was already mentioned, the suitable usage of machine learning techniques in recognizing cybersecurity attacks having logged interaction sessions is to try to learn known normal behavior. Also behavioral patterns are shown to be an important insight into understanding the possibilities of attacks, especially when it is internal usage of a system \citep{greitzer2011modeling, ussath2017identifying, pannell2010user}. Previously multiple approaches for identifying anomalies in interaction sessions were proposed: based on such handcrafted features as length of the session, character distribution in the session, etc. \citep{nascimento2011anomaly, kruegel2003anomaly} 

Behavior inside of a system usually can be described by sequences of particular actions that are happening during the interaction. There are several approaches in machine learning for modeling sequential data, e.g. Hidden Markov Models \citep{yeung2003host}. Here we apply recurrent neural networks, that were also employed in cybersecurity tasks \citep{ishitaki2017application}. Recurrent Neural Networks (RNNs) \citep{Jain:1999} are a class of artificial neural network where neurons are stacked in a \emph{recursive} manner. This allows the network to memorize things from the past by feeding them back to itself. 
Long Short-Term Memory (LSTM) \citep{Hochreiter:1997} is one of the prominent RNN models and has been shown to be capable of avoiding some of the problems which arise during the training of Recurrent Neural Networks.

One of the ways to perform the sequential analysis of interaction sessions is using the tools developed in the field of natural language processing \citep{an2017behavioral, tuor2018recurrent, kim2016lstm}. RNNs, especially LSTMs, are widely used for language modeling \citep{Bengio:2003, MikolovKBCK10, Sundermeyer2012LSTMNN}. The use of RNNs for language modeling dates back to 2003 \citep{Bengio:2003}, and still today some of the recent language models use RNN based networks to produce state-of-the-art results. For example, the LSTM architecture proposed by \citet{wu2016google} produces state-of-the-art results for Neural Machine Translation based on language modeling. 

Applying language models to model users behavior in form of sequences of actions is a straightforward idea since the task of a language model is to predict the next word in a sequence and it is usually trained on a vast amount of unlabeled data. 
Neural language models are language models that utilize the neural network properties to predict the probability of the next word in a sequence \citep{Chen:1996}. In our case, such kind of neural language model is modified, the sequence of user actions is considered to be the sequence of words, thus enabling to model the probability distribution over user actions sequence space.

Language modeling was already employed in the network security community. \citet{tuor2018recurrent} applied character level language modeling for separate lines of logging files in order to identify fraudulent actions. This approach was shown to perform good on one of the publicly available datasets, nevertheless it is rather sensitive to the format of logging information, e.g., having information about the IP address of the request and success of the performed action. It also does not employ the information of the sequences of actions performed in each interaction. Another technique by \citet{kim2016lstm} employs an ensemble of language models that are learning normal users' behavior from sequences of actions. According to \citet{xin2018machine} their approach is one of the most successful in solving the task for one of the publicly available datasets.

Nevertheless, even though both \citet{tuor2018recurrent} and \citet{kim2016lstm} mention the need of separate modeling for particular groups of behaviors, they cover this idea either by ensemble of different models or by specifying the timespan of the sessions to model. A natural way in the context of modeling users' behavior is to cluster sequences based on the user profile. But this information is not always available and might not even exist in some applications, e.g., peer-to-peer interaction where various users might be in control of the node. Moreover, behavioral modeling with recurrent neural networks can be applied in various tasks, e.g. identifying behavior of malware files \citep{rhode2018early}, and binding the approach to users' profiles is a restriction. As it was mentioned by \citet{sommer2010outside} understanding of the exact system and types of interactions is very critical in cybersecurity tasks. For example, \citet{chandrasekhar2014confederation} achieved good results by having an expert in the loop. Based on this, we propose to integrate an interactive visual tool \citep{siming2019lda} for incorporating security experts knowledge about the system. Using a visual interface the experts can separate sequences of interactions to meaningful clusters. The core idea of the solution is to perform multiple runs of LDA topic modeling \cite{blei2003latent}, treating each session $s$ as a document composed of words--actions $a_i$. We run LDA with different parameters, e.g. number of topics, multiple times and get the ensemble of LDA. We use the topics and two matrices (the topic-action matrix and the document-topic matrix) as the input for the visualization system. 

\begin{figure}[!htb] 
  \centering 
  \includegraphics[width=1.0\columnwidth]{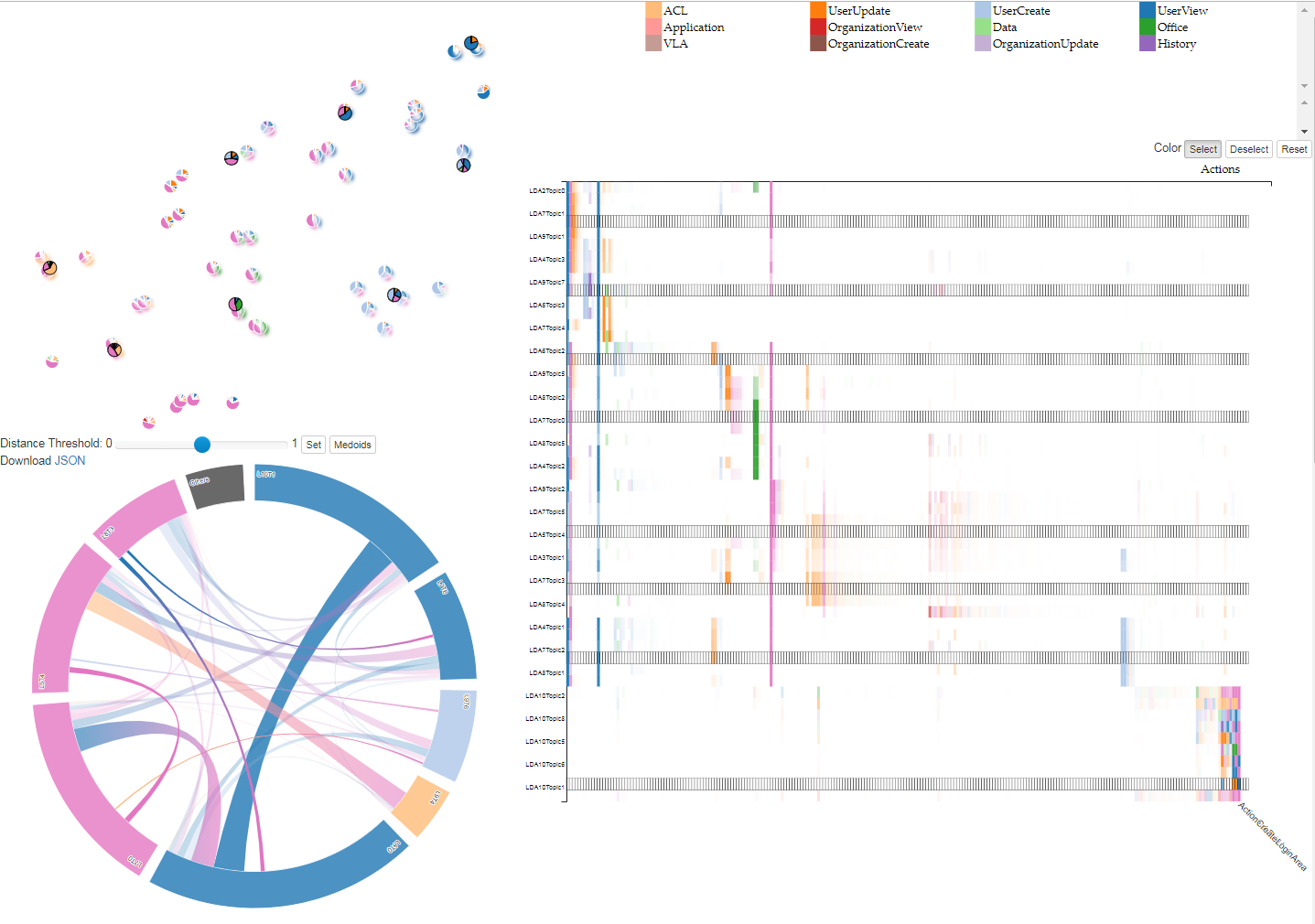} 
\caption{Exemplary view of the visual interface for the security experts for understanding the distribution of actions and exploring created clusters \citep{siming2019lda}.} 
  \label{fig:st_comparison} 
\end{figure}

An exemplary view of the visual interface is shown in Figure~\ref{fig:st_comparison}. It consists of three main parts: topic projection view (top left), topic-action matrix (right) and a topic chord diagram (bottom left). Topic projection view employs the t-SNE technique to visualize the similarity of different topics. It also allows the experts to select or brush the clusters of topics in order to examine the details in the other views. The topic-action matrix view visualizes the probability distribution of actions in each of the topics. The x-axis represents actions and the y-axis represents topics. The higher the opacity of the matrix block, the higher the probability that the action appears in the specific topic. The chord visualization represents the selected topics relationship. The outer fans represent topics, the length of each indicates the number of actions belonging to it. The more actions two topics share, the thicker the links connecting corresponding fans. It is used to indicate how similar topics are.

The result of visual analytics is presented in a form of sets of sessions, so it does not provide an inference technique for a new sequence of actions \cite{siming2019lda}. We have to define a method recognizing the cluster to which the sequence is related the most. There are various approaches for performing this, e.g., simply finding the closest mean to a new sequence or K nearest neighbors. We preferred an approach that allows generalization and comparatively fast prediction--one class support vector machine (OC-SVM) \citep{scholkopf2000support}. OC-SVMs are widely used for anomaly detection and were also applied for intrusion detection \citep{limthong2013real, an2017behavioral}. We employ them only for detecting the corresponding cluster for a new session.

In the next section we describe our approach in more details.

\section{Approach}
\label{approach}
\begin{figure}[!htb] 
  \centering 
  \includegraphics[width=1.0\columnwidth]{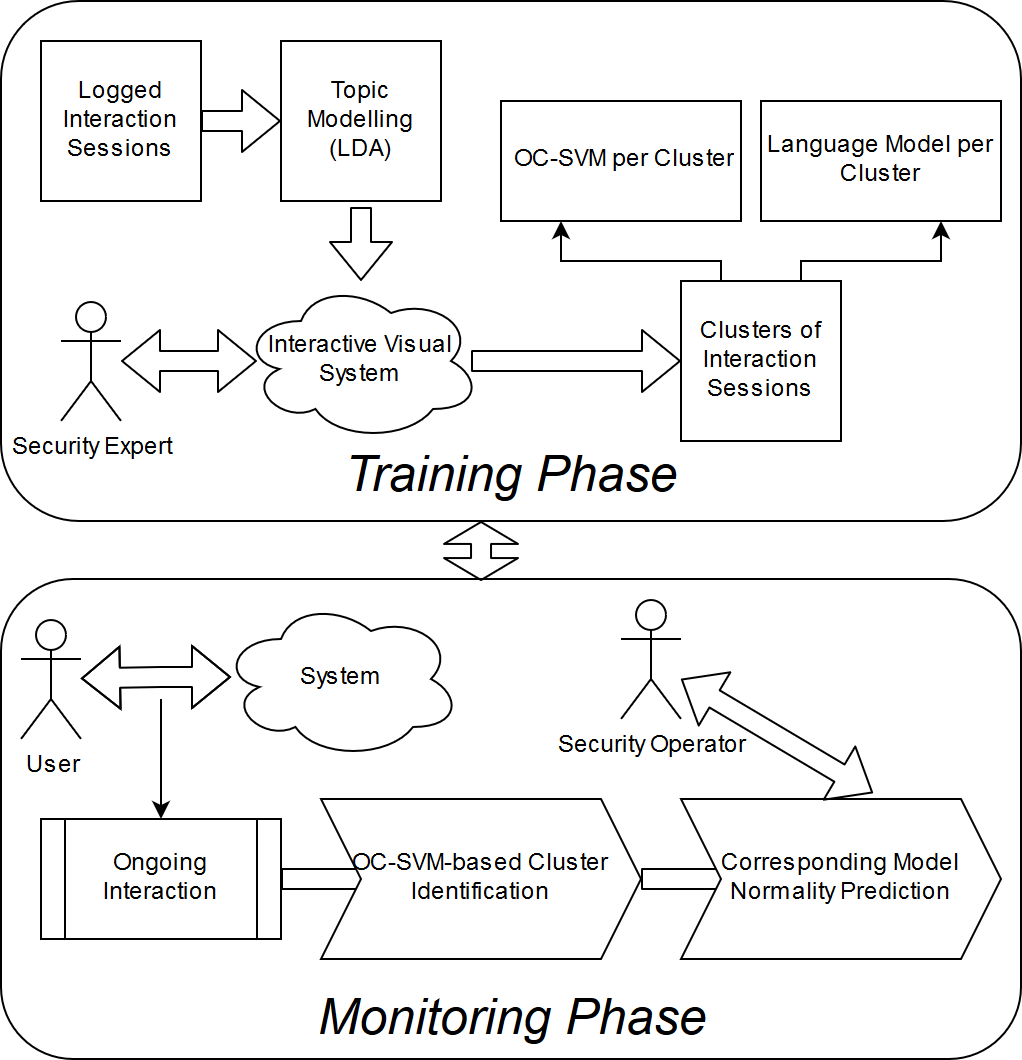}
\caption{Diagram of the proposed approach. The training phase can be repeated at any moment if security experts notice sufficient drift in behavior in the system.} 
   \vspace{-0.1in}   
  \label{fig:approach} 
\end{figure}

Overall, the envisioned pipeline of the approach can be summarized as a diagram depicted in Figure~\ref{fig:approach}. In the following we describe separate parts in more details.

A system is identified by a set of $d\in\mathbb{N}$ possible actions $\mathcal{A}$. Each user interacts with the system using the actions in $\mathcal{A}$. These interactions can be separated into sessions (e.g., all actions between a log-in and a log-out of the system are a session) and such sessions are logged for further investigation. A session of length $n\in\mathbb{N}$ is a tuple $s=(a_1,\dots,a_n)$ with $a_i\in\mathcal{A}$.
In order to use the interactive visual interface to incorporate security experts' knowledge about the system, we perform topic modeling on historical data of $m\in\mathbb{N}$ normal behavior sessions $\mathcal{H} = \{s_i,\dots,s_m\}$. The resulting topics are loaded into the visualization system. 
The security experts can select a group of LDA topics and the medoid is highlighted for further investigation. Topics can be added or removed based on the experts' judgment on whether they are representative or not. 
By investigating the representativeness and coverage of the whole dataset, the experts incorporate their knowledge into the finally selected topics that divide the data $\mathcal{H}$ into clusters for further modeling. As a result of this interaction we obtain $k\in\mathbb{N}$ clusters $\mathcal{G}_i$ such that $\bigcup_{i=1}^{k} \mathcal{G}_i = \mathcal{H}$.

After identifying the clusters of behavior, we train an OC-SVM for each one. Formally, each OC-SVM describes a function $f_i: \mathcal{A}^n \xrightarrow{} \mathbb{R}$ that predicts a score $w_i$ of the cluster $\mathcal{G}_i$ for the input session $s$. During the prediction phase, we compare the scores $w_1,\dots, w_k$ predicted by the OC-SVMs, finding maximal $w^{max}$ denoting the matching cluster $\mathcal{G}^{max}$.

We employ LSTM-based language models for learning behavioral patterns. A separate model is trained on each of the clusters $\mathcal{G}_i$. A language model assigns probability values to sequences of words, analogously we want to assign probability to a sequence of actions. Thus, if a model is trained on normal behavior interactions suspicious behavior will have a low probability. Formally, given an interaction of a user with the system, the goal is to predict the likelihood of an action $a_i$ given the previously observed actions in the session $a_1,\dots,a_{i-1}$. A language model takes observed actions $a_1,\dots,a_{i-1}$ as an input and predicts the probability distribution $\mathbf{p} = \{p_1,\dots,p_d\}$ over all the actions in $\mathcal{A}$. The largest probability $\mathbf{p}^{max}$ is considered to correspond to the next action $a_i$ according to the model. For estimation of the normality of a session $s=(a_1,\dots,a_n)$ we use the average probability of the actually observed actions $a_i\in s$, i.e, $\frac{1}{n} \sum_{i=1}^{n} \mathbf{p}_{a_i}$. The prediction is performed with the model corresponding to the cluster $\mathcal{G}^{max}$ identified with OC-SVM. \citet{kim2016lstm} employed the cross-entropy loss $-\sum_{j=1}^{d} y_j log(p_j)$, where $y$ is a probability distribution with $1$ on the index $j$ of the right action, as an identification of an unlikely action (i.e., low loss means that action is aligned with the behavior learned by the model, while high loss can be interpreted as a warning to further check the session). We are considering the average loss over all the actions $a_i \in s$ in our experiments as well.

Next section describes our experiments setup and shows the results of evaluation.
\section{Evaluation}
\label{evaluation}
In the following we describe the experiments performed on the data provided by the corporate partner in the DiSIEM project. The evaluation is conducted with the aim of both better understanding the system and particular properties of interactions in it as well as to justify generality of the proposed approach. The dataset does not include known misuses of the system, so we validate our approach through (i) checking the ability to model the existing behavior, (ii) creating random sequences of actions and checking their normality, and (iii) evaluating most suspicious sessions with the system experts.

\subsection{Preparatory Evaluation}
\label{preparatory}
\begin{figure}[!htb] 
  \centering 
  \includegraphics[width=0.8\columnwidth]{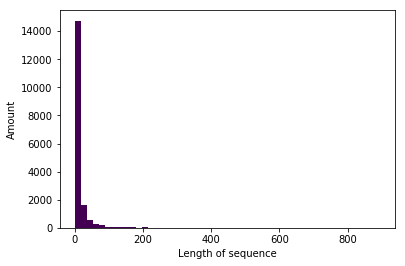} 
\caption{Lengths distribution of the sessions. The longest session consists of more than $800$ actions, while average length is $15$. } 
   \vspace{-0.1in}   
  \label{fig:sessions_lengths} 
\end{figure}

In order to better understand the system that produces the logs we first give insights into the considered dataset. The dataset was recorded for $31$ days and consists of approximately $15000$ sessions performed by $1400$ users with almost $300$ different actions \citep{nguyen2018understanding}. Each session is a sequence of actions, such as 'ActionSearchUser', 'ActionDisplayUser', etc. Since we employ language modeling in our approach, the important information is the lengths of the sequences. The histogram of the lengths distribution can be seen in Figure~\ref{fig:sessions_lengths}. There is some amount of long sessions, but it can be directly seen that most of the sessions is less than $200$ actions long. Indeed, the percentile calculation has shown that $98\%$ of sessions include less than $91$ actions and the average length of a session is $15$.

This analysis allows to conclude that we can employ minibatch training for the language models (i.e., presenting several examples to the network at one step of optimization) which is an efficient way to achieve a faster convergence. The chosen length of input sequences is $100$, so more than $98\%$ of sessions can be learned fully. Each session then is presented as batch of training data via a moving window of length $100$, i.e. first element of batch is filled with zeros in the beginning and first action of the session in the end, while the last element of batch includes all the actions of the session till the last. We also eliminate the sessions consisting of less than two actions, because then there is no observed and predicted part for the model to learn.

Next step of the data transformation is to represent it in the form that is suitable for the input to a LSTM network. 
We use one-hot-encoding technique for it: every action is described as a vector of length $d$ in the form $a_i := [e_1,\dots,e_d]$ where $e_j=1$ iff $j==i$ and otherwise $e_j=0$. After padding and cropping each sequence of actions in the dataset is represented by a two-dimensional matrix $100 \times d$. Every example is treated like a sequence that should be continued, i.e., the model has to predict which action follows it. Thus, an input example is a $99$-actions long sequence and output is the $100$th action.

Learning a neural network requires identifying multiple hyper parameters, such as amount of neurons per layer, minibatch size most suitable for training, etc. We have chosen an architecture of LSTM network, which consists of one layer with LSTM units, a dropout layer \citep{srivastava2014dropout} following it, and a final dense layer with softmax activation for the prediction of the next action in a sequence via probability distribution over all possible actions. 
The evaluation was performed on a small subset of the data and the final configuration looks as following: $256$ LSTM units followed by a dropout layer with $0.4$ dropout rate should be trained with a minibatch size of $32$ and a learning rate of $0.001$. Since we considered the full dataset for evaluation of hyper parameters it might happen that additional reevaluation for each of the clusters can improve the results. Nevertheless, this is left for the future exploration.

\subsection{Clusters Identification}
As described in Section~\ref{approach}, multiple topic models are learned on the sequences of users' actions and the results are loaded into the visual interactive system for security experts analysis. This analysis on the considered dataset was performed by the experts from the corporation owner of the data. As a result, $13$ clusters were identified, each carrying particular semantic meaning. We performed frequent patterns mining for the discovered clusters and found out that, for example, one of them includes all the sessions with actions to unlock user's access to the system, another includes all modifications of roles of users, third has all the actions concerned with edition of office entities. That corresponds to the intent to identify particular behavior clusters. For each of the discovered clusters corresponding sessions were organized into datasets, each dataset split into training, validation and testing parts (in proportion $70/15/15$). After modifications described in Subsection~\ref{preparatory} both LSTM language models with pre-evaluated hyper parameters and OC-SVMs were trained on each of them. 

\begin{figure}[!htb] 
  \centering 
  \includegraphics[width=0.8\columnwidth]{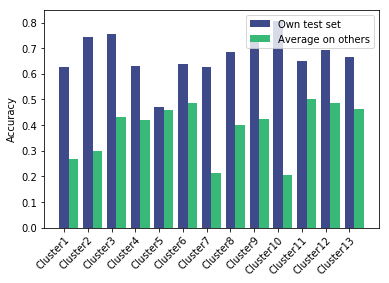}
\caption{Comparison of the test accuracy of cluster models calculated on the corresponding testing set against the average accuracy of the same model on all the other testing sets.} 
   \vspace{-0.1in}   
  \label{fig:one_vs_avg} 
\end{figure}

In order to validate the diversity of the obtained models we perform a simple test demonstrating the performance of the cluster models. We calculate the accuracy (i.e., percentage of correctly predicted actions) for each of the models first on the testing set of the corresponding cluster and then on all the others. The result is depicted in Figure~\ref{fig:one_vs_avg}. The clusters for this experiment were ordered by their size in ascending order. First conclusion that can be made from this plot is that larger clusters produce much stronger models, e.g., performing good on any cluster data, but nevertheless even very small clusters (the smallest includes only $177$ sessions) learn the prediction task. Second conclusion supports the diversity of the models--even the strongest models perform much better on their own testing set than on average on the testing sets of all the other clusters. Thus, we obtained specific models for each of the semantically meaningfull clusters of behaviors.

After verifying the performance of the cluster models, our next step is to compare them to baselines. As a very strong baseline we selected a model trained on the whole dataset. Comparing to this baseline, one should keep in mind that overall in real datasets the amount of interaction sessions is constantly growing--thus clusters can be made arbitrary large making the corresponding models much stronger. And as it was already grounded in Section~\ref{approach}, global models in cybersecurity with high probability will suffer from less accurate fraud detection. The second baseline is a global model trained on an arbitrary subset of the data of the same size as the cluster dataset.

\begin{figure}[!htb] 
  \centering 
  \includegraphics[width=0.8\columnwidth]{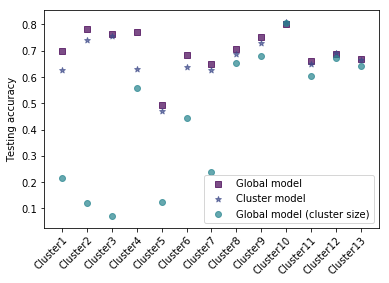} 
\caption{Clusters are organized in ascending order by size. The accuracy achieved on the individual testing sets by the cluster model is compared to the accuracy of the global model and global model trained on the subset of the size equal to the cluster size.} 
   \vspace{-0.2in}   
  \label{fig:acc_compare} 
\end{figure}

We analyze accuracy achieved on the testing sets for each of the clusters. The result is plotted in Figure~\ref{fig:acc_compare}. The global model (trained on cluster size data subset) baseline directly shows that the knowledgeable identification of clusters is extremely important for the approach. The accuracy of the models trained on the arbitrary clusters of the same size cannot compete with the performance of the cluster models when the size of data is not large enough. Concerning the strong baseline we can see the larger gap in performance for small clusters, but as soon as the size is becoming sufficient for proper training of the model, the performance is becoming as good or even better (no matter that the largest cluster has only around $3500$ examples compared to more than $10000$ sequences for the global model). A similar result in terms of loss values can be found in Appendix~\ref{appendix}.

\subsection{Use Case: Online Regime}
Having evaluated the performance of cluster models against baselines, we move to the realtime use case. We want to see how our approach works in the online regime, i.e., when the session is analyzed in realtime, action by action in order to give an alarm for security operators as soon as some suspicious behavior is observed. For this we unite all the testing datasets of the clusters and present it to the OC-SVMs and corresponding models action by action. In the previous experiments we were assuming that we know the cluster of each session, while here we include OC-SVM prediction score for identifying the most matching cluster.

\begin{figure}[!htb] 
  \centering 
  \includegraphics[width=0.8\columnwidth]{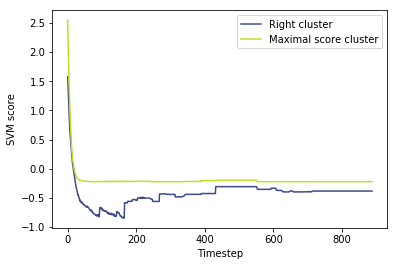} 
\caption{Development of scores predicted by OC-SVMs per action. We compare the score predicted by the \emph{right} OC-SVM, i.e., corresponding to the cluster that the session really belongs to, against the maximal score among all the OC-SVMs.} 
  \label{fig:osvm_preds} 
\end{figure}

One of the insightful results of this experiment is concerned with the predictions of OC-SVMs. The per action prediction scores, averaged among all the testing sessions are shown in Figure~\ref{fig:osvm_preds}. We checked both the score of the OC-SVM that should give the largest value, i.e., the OC-SVM corresponding to the cluster of the session and the maximal achieved score among all of them. One can directly see from the development of scores an indication that all the sessions longer than the average length are considered to be outliers by all the OC-SVMs. This conclusion is supported by the wide use of the OC-SVMs for outliers detection. In order to fight the consequences of such behavior we propose to check the cluster only during first $15$ actions (as it was mentioned before, average length of the session is $15$) and then use the most frequently assigned cluster as the predicted one for the session.

\begin{figure}[!htb] 
  \centering 
  \includegraphics[width=0.8\columnwidth]{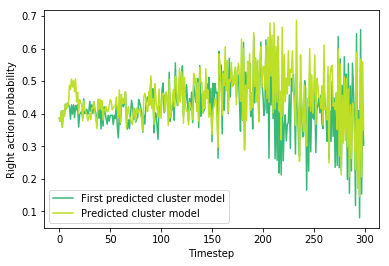}
\caption{Online regime of approach application. Average of likelihood for each next action in each of the testing sessions is calculated for two baselines: predicted on every step model, and predicted during first $15$ actions model.} 
   \vspace{-0.1in}   
  \label{fig:prob_pred_per_timestep} 
\end{figure} 

As a result of the use case we show \emph{scores} development per action performed during the session. By score here we mean the probability of the action that really happened according to the model prediction. The scores for all the testing sessions are averaged. The plot for the two baselines is shown in Figure~\ref{fig:prob_pred_per_timestep}: 
\begin{enumerate}
    \item the predicted cluster model from the maximal OC-SVM score,
    \item the predicted cluster model from the first $15$ predictions of OC-SVM scores.
\end{enumerate}
These baselines were selected as the ones that can be realistically used in real application. We restrict the sequence length considered in the plot to $300$ actions. We can observe that the level of likelihood is rather stable for the first $100$ actions, but further on it decreases while variance increases considerably. We also can see that proposed identification of cluster during first actions leads to more stable development of the scores, without significant drop in the beginning. 

Thus, the use case indicated the usefulness of our approach for online monitoring of interactions--as soon as predictions start vary a lot or drop down considerably that is the alarm to the security operator.

\subsection{Evaluation of Normality Prediction}
Finally we are evaluating how good our approach is capable of identifying normality of an actions sequence. For this we consider the average likelihood of each action in the session as a normality measure of the session, as it was introduced in Section~\ref{approach}. We also consider average loss across a session to estimate this normality measure as well. Here we summarize the results across the clusters, separately calculated estimations can be found in Appendix~\ref{appendix}.

In our dataset we do not have sessions showing malicious behavior, so in order to have a rough estimation of possible results on outlying sessions we introduce an artificial test set. This test set contains same amount of sessions as the main data test set, each session has a randomly chosen length in an interval $[5, 25]$ and each action is randomly chosen from the set of actions $\mathcal{A}$. In order to evaluate normality of these sessions, we employed our prediction pipeline.

\begin{figure}[!htb] 
  \centering 
  \includegraphics[width=0.6\columnwidth]{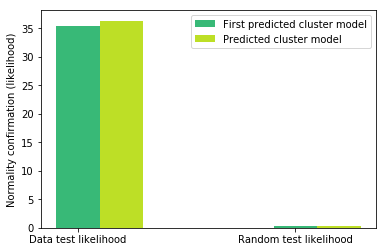}
  \caption{Normality estimation in terms of likelihood and average loss suffered at each action in the session.} 
  \label{fig:norm_scores_art} 
\end{figure}

\begin{figure}[!htb] 
  \centering 
  \includegraphics[width=0.6\columnwidth]{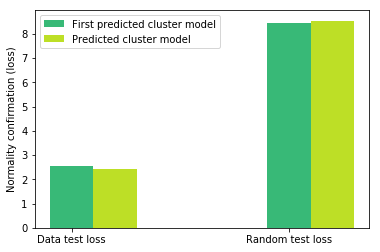}
  \caption{Normality estimation in terms of likelihood and average loss suffered at each action in the session.} 
   \vspace{-0.1in}   
  \label{fig:norm_loss_art} 
\end{figure} 

Figure~\ref{fig:norm_scores_art} shows estimations made with average likelihood, while Figure~\ref{fig:norm_loss_art} shows average loss. As desired, the average likelihood on the artificially generated test set is extremely low (on the level of random prediction) and the average loss on this dataset is almost twice higher than the average loss on the real testing data. The difference in likelihood estimations of normality is much more drastic than in terms of average loss. But both of the metrics allow to directly distinguish between the true test set and abnormal one.

Finally, we also presented the most suspicious according to our approach sessions to the system experts. According to their specifications, such actions as 'ActionUnLockDisplayedUser', 'ActionResetPwdUnlock' or 'ActionDeleteUser', i.e., active modifications of existing user profiles, are most alarming. Among top $20$ sessions we found for example the following: 'ActionSearchUsr', 'ActionWarningDeleteUser', 'ActionDeleteUser', 'ActionCreateUser', 'ActionCreateUser', 'ActionCreateUser', 'ActionCreateUser', 'ActionSearchUsr', 'ActionUnLockUser', 'ActionCreateUser', 'ActionSearchOffice', 'ActionDisplayOneOffice', 'ActionDisplayDirectTFARule'. Such sessions are exactly the ones that should give alarm notification to the operators.

\section{Conclusion}
\label{conclusion}
In our research we addressed the task of modeling behavior during interactions with a system. We proposed a complex approach consisting of identification of clusters of possible behavior with involvement of the system security experts and application of language modeling with LSTM networks for learning the normal behavior. This modeling can afterwards be employed to identify suspicious outlying sessions.

We evaluated our approach on a dataset provided by our partner organization in the project DiSIEM. Since the dataset contains only examples of normal behavior our main goal was to model it. In the future we want to consider one of the publicly available datasets (such as ADFA \citep{creech2013generation}) in order to compare our approach to the others \citep{xin2018machine} and evaluate its ability for identifying malicious behavior.

We can point out possible future directions for improving and developing the proposed approach. First, weighted combination of multiple scores from cluster models might give more objective score, taking into account possible imprecision of cluster identification. Second, identification of trends in the development of the scores in order to set the alarm for security operators can perform better than reacting to every low score right away. Third, perplexity score might be more objective normality measure of a session than the average per action loss or likelihood.

\newpage

\bibliographystyle{IEEEtranN} 
\bibliography{bibliography}

\clearpage

\section{Appendix}
\label{appendix}
\begin{figure}[!htb] 
  \centering 
  \includegraphics[width=0.8\columnwidth]{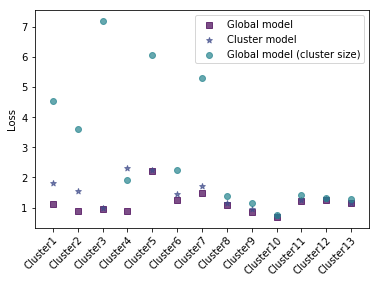} 
\caption{Clusters are organized in ascending order by size. The loss achieved on the individual testing sets by the cluster model is compared to the loss of the global model and global model trained on the subset of the size equal to the cluster size.} 
  \label{fig:losses_compare} 
\end{figure}

Per cluster evaluation in terms of loss values achieved by models is shown in Figure~\ref{fig:losses_compare}. 

\begin{figure}[!htb] 
  \centering 
  \includegraphics[width=1\columnwidth]{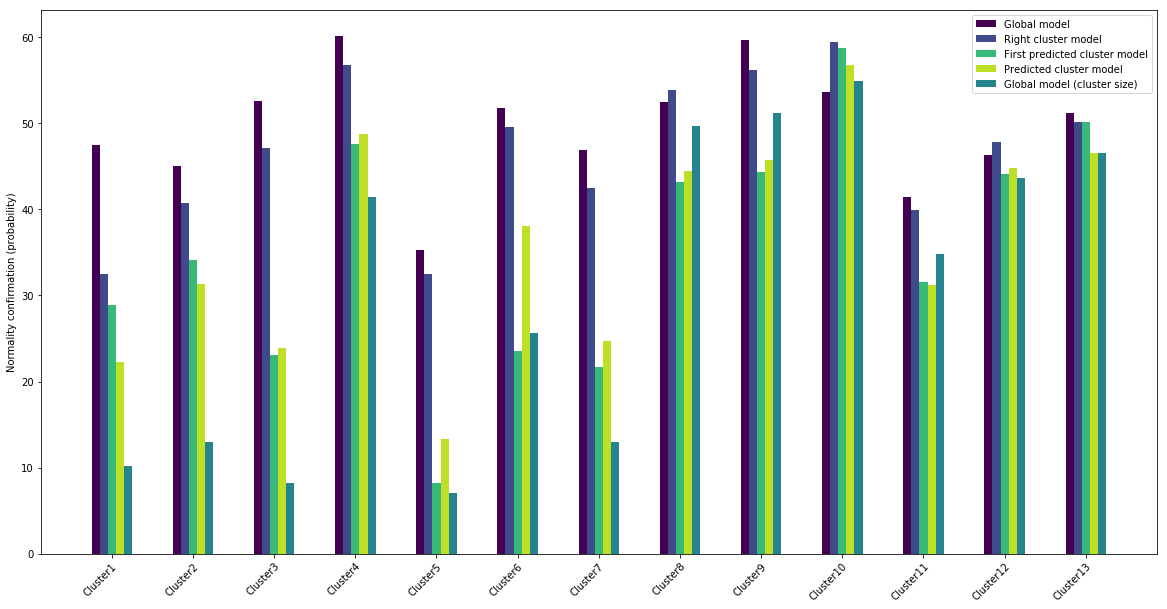}
    \caption{Normality estimation in terms of average likelihood of each action in the session.} 
  \label{fig:norm_prob} 
\end{figure}

Normality estimation of sessions was performed on the test dataset for all the four considered previously baselines. The result is depicted in Figure~\ref{fig:norm_prob}. We again can observe higher normality scores for the stronger models trained on larger clusters. Overall we can see that the identification of the cluster model performs sufficiently well, and identification of the cluster based on the first actions allows to avoid consequences of OC-SVMs peculiarities. 

\begin{figure}[!htb] 
  \centering 
  \includegraphics[width=1\columnwidth]{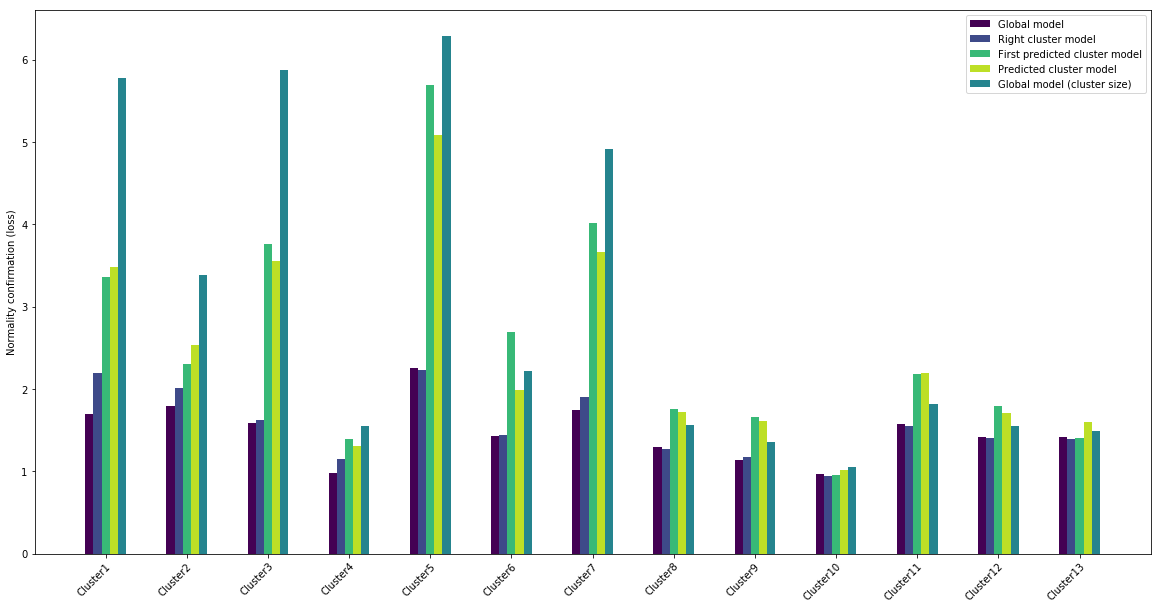}
\caption{Normality estimation in terms of average loss suffered at each action in the session.} 
  \label{fig:norm_loss} 
\end{figure}

Other than average likelihood we considered average loss suffered during each action prediction, following \citet{kim2016lstm}. The results are shown in Figure~\ref{fig:norm_loss} and follow same pattern as for average likelihood.

\end{document}